\documentstyle[12pt]{article}

\input{epsf.tex}

 \newcommand {\Stretch}[1]{
\renewcommand{\baselinestretch}{#1 }\large\normalsize}
 
\newcommand {\bm}[1]{{\mbox{\boldmath $#1$}}}

\def\beq{\begin{equation}}
\def\eeq{\end{equation}}
\def\bea{\begin{eqnarray}}
\def\eea{\end{eqnarray}}

\def\ag{A_g^+}
\def\bu{B_{u}^-}

\def\s0{\phi_0}

\def\n2{\nabla^2}

\begin{document}

\Stretch{3}

\setcounter{page}{1}
\baselineskip=25pt \parskip=0pt plus2pt

\begin{center}
{\large \bf 
A Theoretical  Investigation of the Low Lying 
Electronic Structure of Poly(p-phenylene vinylene).}

\Stretch{1.5}

{
Mikhail Yu.\ Lavrentiev$^{1*}$, William  Barford$^1$, 
Simon J.\ Martin$^1$, Helen Daly$^1$ and
Robert  J.\  Bursill$^2$
 \\ }
$^1$Department of Physics and Astronomy,
The University of Sheffield,
Sheffield, \\ S3 7RH, United Kingdom \\
$^2$School of Physics, University of New South Wales, Sydney,
NSW 2052, Australia\\

\end{center}
\Stretch{1}

\begin{abstract}

The two-state molecular orbital model of the one-dimensional phenyl-based
semiconductors is applied to poly({\em para}-phenylene vinylene).
The energies of the
low-lying excited states are calculated using the
density matrix renormalization group method. Calculations of both the
exciton size and the charge gap show that there are both $^1\bu$ and
$^1\ag$ excitonic levels below the band threshold. 
The energy of the $1^1\bu$ exciton extrapolates to 2.60 eV in the limit
of infinite polymers, while the energy of the $2^1\ag$ exciton extrapolates
to 2.94 eV.
The calculated binding energy of the $1^1\bu$ exciton is 0.9 eV
for a 13 phenylene unit chain and 0.6 eV for an infinite polymer.
This is expected 
to decrease  due to solvation effects. 
The lowest triplet state is calculated to be at ca.\ 1.6 eV, with the
triplet-triplet gap being ca.\ 1.6 eV.
A comparison between theory, and
two-photon absorption and electroabsorption is made,
leading to a consistent picture of the 
essential states responsible for most of the third-order
nonlinear optical properties.  An interpretation of the experimental
nonlinear
optical spectroscopies suggests an energy difference of ca.\ 0.4 eV between
the vertical
energy  and ca.\ 0.8 eV between the relaxed energy, of the $1^1\bu$ exciton
and the band gap, respectively.

\end{abstract}


\section{Introduction}

Since the discovery of the electro-luminescent properties 
of the organic semiconductor poly(para-phenylene vinylene) (PPV) 
\cite{nature90} an understanding of its low lying electronic 
structure has remained a challenge.   The observation of electroluminescence
implies that there is a one-photon transition from the lowest excited 
singlet state to the ground state.  The structure of PPV, a sequence
of phenylene and vinylene units capped by phenyl rings, is shown in Fig. 1. 
Since PPV is centro-symmetric, and thus possess $C_2$ symmetry, the ground
state is spatially symmetric ($A_g$), while the first excited singlet 
state is therefore odd under rotation ($B_u$).  Furthermore, PPV, 
to a very good approximation, is non-polar (i.e.\ each atom is charge 
neutral) and thus another symmetry it possesses is particle-hole 
symmetry.  As one-photon transitions occur between states of opposite 
particle-hole symmetry, we will confine out attention to the $\ag$  and 
$\bu$ symmetry sectors in this paper.

\begin{figure}[htbp]
\caption{
Poly(para phenylene vinylene). $N=0$ corresponds to stilbene.
}
\centerline{\epsfxsize=10.0cm \epsfbox{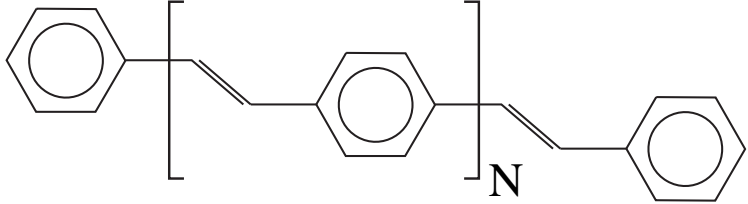}}
\end{figure}

A number of optical probes 
have been employed to ascertain the character of the low lying 
states.  In particular, one-photon absorption identifies the 
$^1\bu$ states, two-photon absorption (TPA)
identifies the $^1\ag$ states, and  
electroabsorption (EA) and third harmonic 
generation (THG) identify both kinds of states.  A consistent picture 
of the low lying
electronic structure of PPV has been slow to emerge, owing to sample 
variability (including the use of PPV derivatives),  inter-chain effects, 
and the fact that some measurements probe vertical transitions, while 
others probe relaxed transitions.
Nonetheless, a consistent experimental interpretation is beginning to 
appear, and we discuss and interpret current experiments in the
light of our and other theoretical calculations.

A full theoretical treatment must  include the effects of the strong 
electron-electron interactions, electron-lattice relaxation and 
inter-chain interactions to fully describe PPV thin films.  So far, 
however, most calculations have considered 
single chains in the absence of electron-lattice relaxation.  Furthermore, 
they have included 
electron-electron interactions in the most simple way, e.g., by 
Hartree-Fock mean field or by the single configuration interaction (SCI), 
neither of which can describe
the highly correlated $^1\ag$ states of conjugated semiconductors.  

A remarkably reliable description of $\pi$ conjugated electron systems 
is provided by the semi-empirical one band Pariser-Parr-Pople (P-P-P) model.  
Both Shimoi and Abe
\cite{shimoi96}, and Chandross and Mazumdar \cite{chandross97} solved a 
re-parametrised P-P-P model, within the SCI approximation, to obtain the 
optical transitions and conduction band edge.  Shimoi and Abe predicted 
a $1^1\bu$ exciton at 2.4 eV and the band edge at 3.2 eV, while Chandross 
and Mazumdar obtained results of 2.7 eV and 3.6 eV for the  $1^1\bu$ 
exciton and band edge, respectively.
Gomes da Costa and Conwell \cite{gomes93} used as their model of PPV 
thin films the physics of three
dimensional semiconductors and Wannier excitons.  
Using a phenomenological dielectric constant they obtained the $1^1\bu$ 
exciton at 2.4 eV and the band gap at 2.8 eV.
Rice and Gartstein \cite{rice94} presented a phenomenological model 
based on the molecular orbitals which was solved analytically.  They 
predicted a binding energy of ca. 0.1 eV.
In a later paper \cite{gartstein95}, a more elaborate model
was presented, with a binding energy lying between 0.2 and 0.4 eV.
Beljonne et al. \cite{beljonne98} performed a INDO/MRD-CI study
of PPV oligomers in a molecular model. By extrapolating to an eight
phenylene
ring oligomer,
they associate the onset of
the conduction band by the $n^1\bu$ state at 4.73 eV, and 
locate other important states below it: $1^1\bu$ at 3.13 eV,
$2^1\ag$ at 3.78 eV, and $m^1\ag$ at 4.28 and 4.73 eV. Thus, their results
suggest that these states are essentially excitonic. 
Harigaya \cite{harigaya97} performed a single-CI study of PPV, 
poly(p-phenylene) (PPP)
and related polymers. In particular, for PPV he found the Hartree-Fock
gap at 3.64 eV and the onset of long-range excitons at only
a slightly lower energy of 3.60 eV.

In this paper we employ the 
two-state molecular orbital (MO) model introduced in 
\cite{barford97} and \cite{barford98} to calculate the  energies and 
correlation functions of the low lying, predominately long
axis polarised, states of single oligomers.  
The model and its parametrisation is discussed in
\S 2.  In this treatment we also neglect electron-lattice relaxation 
and inter-chain effects.  However, 
since the model is solved by the density matrix renormalisation group 
(DMRG) method \cite{white93}, an essentially exact treatment of one
dimensional quantum 
Hamiltonians, electron-electron
correlations are correctly treated.  As discussed in \S 3, we predict 
bands of $\bu$ and $\ag$ excitons.  We identify the 
onset of the unbound states by an examination of the particle-hole
correlation function.  We further show that the first 
unbound $^1\ag$ state couples strongly to the $1^1\bu$ exciton, 
and hence contributes to the EA spectrum.  
In \S 4 the TPA and EA spectra 
of oligophenylene-vinylenes are calculated.
By comparing the calculated and experimental 
TPA and EA spectra, we deduce a coherent picture of the energies and 
symmetries of the low lying states.  We conclude in \S 5.

\section{The Two State Model}

The essential assumption underlying the two-state MO model is that the six 
MOs of the phenyl(ene)
 ring, arising from the six conjugated $\pi$ electrons,
may be replaced by the bonding HOMO and  LUMO states. 
In principle, the many body correlations between the MOs can be determined 
from the underlying P-P-P model.  When this was done for 
PPP, however, it was found that there is a poor agreement with 
experiment \cite{bursill98b}.  This is a result of the neglect of the 
many body correlations involving the neglected orbitals.  Nevertheless, 
if a description of the low energy predominately long-axis polarised 
states is required, a two-state model should contain the essential 
physics. (The meaning of a predominately long-axis polarised state 
will be discussed shortly.)   Thus the two-state model is a 
{\em phenomenological} model whose parameters are chosen to agree 
with exact P-P-P model calculations of benzene, biphenyl and stilbene.   
This approach was adopted for  PPP in \cite{barford98}, and where 
comparison to experiment could be made, was found to be succesful.  We 
will show that its application to PPV also leads to a consistent 
agreement with experiment. 
The transformation from atomic to molecular orbitals
is performed for phenyl(ene) and vinylene units, and below we 
denote the HOMO orbital by $\left|1\right\rangle$ and the LUMO by 
$\left|2\right\rangle$ for both types of molecular units. 

The 2-MO model reads,
\begin{eqnarray}
H & & = -\sum_{i\,\alpha\,\beta\,\sigma} t_{\alpha\beta}
 \left[ a_{i\alpha\sigma}^{\dagger} a_{i+1\beta\sigma}
 + \rm{ h.c.}\right] + \sum_{i\,\alpha}
\epsilon_{\alpha}(n_{i\alpha}-1) + U \sum_{i\,\alpha}
 \left(n_{i\alpha\uparrow}-
\frac{1}{2}\right) \left(n_{i\alpha\downarrow}-
 \frac{1}{2}\right) \nonumber \\
& &
+\;\frac{U}{2} \sum_{i\,\alpha\neq\beta} (n_{i\alpha}-1)
 (n_{i\beta}-1) + V \sum_{i \,\alpha\,\beta} (n_{i\alpha}-1)
 (n_{i+1\beta}-1) \nonumber \\
& &
-\;X \sum_{i\,\alpha\neq\beta}
\left[ \bm{S}_{i\alpha}.\bm{S}_{i\beta} +\frac{1}{4}
\left(n_{i\alpha}-1\right)
\left(n_{i\beta}-1\right) \right]
\nonumber\\
& &
+\;\frac{P}{2} \sum_{i\,\alpha\neq\beta\,\sigma}
 a_{i\alpha\sigma}^{\dagger}
a_{i\alpha\bar{\sigma}}^{\dagger} a_{i\beta\bar{\sigma}}
a_{i\beta\sigma},
\label{model}
\end{eqnarray}
where $a_{i\alpha\sigma}^{\dagger}$ creates an electron in the
molecular orbital $\alpha$ on the molecular repeat
unit $i$ with the spin $\sigma$;
$\bm{S}_{i\alpha}=\sum_{\rho \rho^\prime}
a_{i\alpha\rho}^{\dagger}
 \bm{\sigma}_{\rho\rho^\prime}a_{i\alpha\rho^\prime}$
and $\bm{\sigma}$ are the Pauli
spin matrices.
The creation of an exciton in this model corresponds to the excitation
of an electron from a HOMO to a LUMO state on the same or a near by unit,
resulting in a bound electron-hole pair.

The intra-phenyl(ene) parameters were derived in \cite{barford98} by 
comparing to P-P-P model calculations of benzene and biphenyl \cite{PPPmodel}.
Since the vinylene unit is exactly described by its HOMO and LUMO 
states, its intra-unit interactions are found by an exact mapping from the 
atomic orbital basis of the P-P-P model to the MO basis.  The P-P-P model 
parameters are those derived in \cite{PPPmodel}.   The remaining parameters 
are the inter-unit Coulomb repulsion, $V$ and the inter-unit 
hybridisation, $t$.  We will assume that
$ V = (U_{phenylene} + U_{vinylene})/4$.  Finally, $t$ is parametrised by 
fitting the 2-MO model prediction of the $1^1\bu$ exciton in stilbene to a 
P-P-P model calculation, which puts it at 4.17 eV \cite{castleton} (in 
excellent agreement with experiment).  The parameters are summarised in 
Table 1.

\begin{table}[htbp]

\centering

\begin{center}
\begin{tabular}{|c|c|c|}
\hline
 Parameter & Phenyl(ene) unit & Vinylene unit\\
\hline
Intra-unit HOMO-LUMO gap & 5.26  & 5.43 \\
\hline
Intra-unit Coulomb repulsion & 3.67  & 8.70  \\
\hline
Inter-unit Coulomb repulsion & 3.09  & 3.09  \\
\hline
Intra-unit exchange energy & 0.89  & 1.36   \\
\hline
Intra-unit pair hopping & 0.89   & 1.36    \\
\hline
Inter-unit hybridisation & 1.27  & 1.27  \\
\hline

\end{tabular}

\end{center}

\caption{ Parameters used in the 2-MO model (eV). }

\end{table}

We can use the P-P-P model calculation of stilbene to evaluate the success
 of the 
2-MO model for other states.  Before doing that we need to discuss the effect
 of 
long range Coulomb correlations on the spatial symmetry of the electronic
 states.  
In the absence of Coulomb interactions the P-P-P Hamiltonian of stilbene (and
 PPV) 
possess topological $D_{2h}$ symmetry.  The irreducible representation of the 
$D_{2h}$ group are $A_g$ (symmetric), $B_{1u}$ (long-axis polarised), $B_{2u}$
(short axis-polarised) and $B_{3g}$ (anti-symmetric).   The effect of the 
Coulomb interactions is to weakly break the topological $D_{2h}$ symmetry and 
render the molecule $C_2$
symmetric.  Thus, the electronic states  evolve from $A_g$ and $B_{3g}$
 to $A_g$, 
and $B_{1u}$ and $B_{2u}$ to $B_u$.  Nevertheless, the dipole
 active states are 
either {\em predominately} long-axis or short-axis polarised.  Now,
 the 2-MO model 
describes $B_u$ states which are predominately long-axis  polarised
 (i.e. $B_{1u}$ 
states
in the absence of symmetry breaking terms) and $A_g$ states which
 would be $A_g$ 
(and not $B_{3g}$) in the absence of symmetry breaking terms.
  The relevant lowest 
excited $A_g$ state of stilbene to compare with the 2-MO model
 prediction is at 
5.18 eV \cite{footnote}.  The 2-MO model value of 5.24 eV has a
 relative error of 
1.3 \%.  Similarly, the P-P-P model value of 2.77 eV for
 the $1^3B_u^+$ state 
compares well with the 2-MO model prediction of 2.65 eV (a relative error of 
-4.2 \%).

The calculations of the energies of the ground and lowest excited states,
as well as their properties
are performed using the DMRG method \cite{white93}. The details of the
implementation of the method are given in \cite{barford98}, together with
results of numerous accuracy tests.

\section{The low energy spectra and correlation functions}

\subsection{The Singlet Spectrum}

Fig.\ 2 (a) shows the energy spectra of 
the key low lying $^1\bu$ 
states as a function of oligomer length. The 
energy of the lowest $^1\bu$ state extrapolates
to 2.60 eV in the limit of infinite polymers.
The  experimental oligomer vertical energies of the $1^1\bu$ exciton,
obtained from one-photon absorption \cite{woo93}, are
plotted in Fig.\ 2(a).  These are in excellent agreement with our
calculation. The typical PPV thin film result of ca.\  2.8
eV for the vertical transition  \cite{martin98} is also in close agreement 
with the infinite polymer result of 2.60 eV.
Note that the $0-0$ transition is typically 2.4 eV  \cite{martin98}.
The low lying $^1\ag$ spectrum is shown in Fig.\ 2(b).
The energy of the first excited  $^1\ag$ state ($2^1\ag$)
extrapolates to
2.94 eV.

\begin{figure}[htbp]
\caption{
(a) The energies of the key lowest bound and  band
states of $^1\bu$ symmetry, and the charge gap,
$E_G$,
as a function of the number of phenylene units.
$1^1\bu$ (solid diamonds), $2^1\bu$ (solid triangles),
$3^1\bu$ (solid squares),
 $n^1\bu$ (open diamonds),
and $E_G$ (dashed line).
The experimental results are shown as  crosses \protect\cite{woo93}.
(b) The energies of the key lowest bound and  band
states of $^1\ag$ symmetry, and the charge gap,
$E_G$,
as a function of the number of phenylene units.
$2^1\ag$ (solid squares),
$3^1\ag$ (solid diamonds),
$4^1\ag$ (solid triangles),
 $m^1\ag$ (open squares),
and $E_G$ (dashed line).
}
\centerline{\epsfxsize=15.0cm \epsfbox{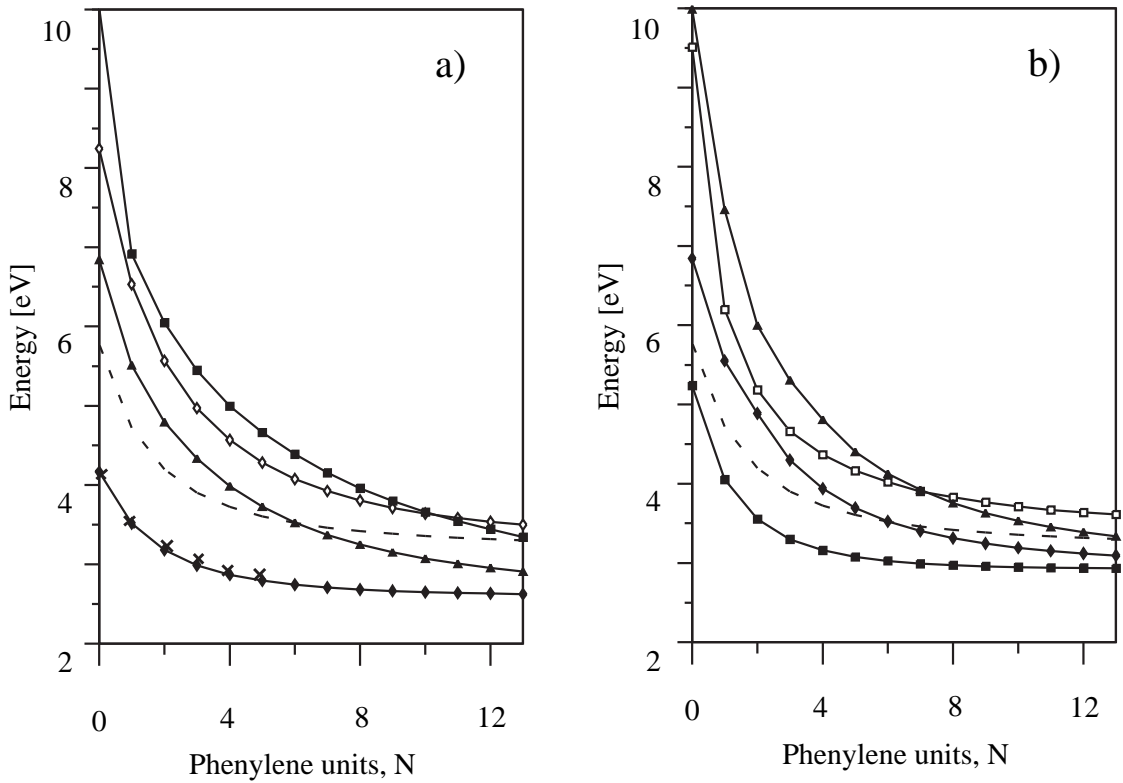}}
\end{figure}
  
To distinguish between bound (i.e.\ exciton) and 
unbound (i.e.\ band) states, the spatial correlation function 
\cite{barford98}, defined as
\begin{eqnarray}
C_{ij}(\left|n\right\rangle) = \left\langle{n}\right|{S_{ij}^{\dagger}}
\left|{1^1A_g^+}\right\rangle,
\label{correlation}
\end{eqnarray}
is calculated. Here, $ S_{ij}^{\dagger} $ is a singlet exciton creation
operator, which removes a
particle from the HOMO on repeat unit $j$ and places it into
the LUMO on repeat unit $i$:
\begin{equation}
S_{ij}^{\dagger} = \frac{1}{\sqrt{2}} (a_{i2\uparrow}^{\dagger}
a_{j1\uparrow} + a_{i2\downarrow}^{\dagger} a_{j1\downarrow})
\label{exciton}
\end{equation}
This correlation function is  used to calculate the mean
particle-hole spacing, as shown in Fig.\ 3.
The particle-hole separation of the $1^1\bu$ and $2^1\ag$ excitons are 
roughly one
and two to three phenylene-vinylene repeat  units, respectively. 
The $^1\bu$ excitons are more tightly bound than the  $^1\ag$ excitons, 
because the former, being negative
under the particle-hole operator, are `s'-wave excitons, while the latter,
being positive
under the particle-hole operator, are `p'-wave excitons \cite{barford98}.

\begin{figure}[htbp]
\caption{
The mean electron-hole distance of the key lowest bound and 
band states as a function of the
number of phenylene units. $1^1\bu$ (solid diamonds),
$2^1\ag$ (solid squares),
 $n^1\bu$ (open diamonds),
and  $m^1\ag$ (open squares).
}
\centerline{\epsfxsize=10.0cm \epsfbox{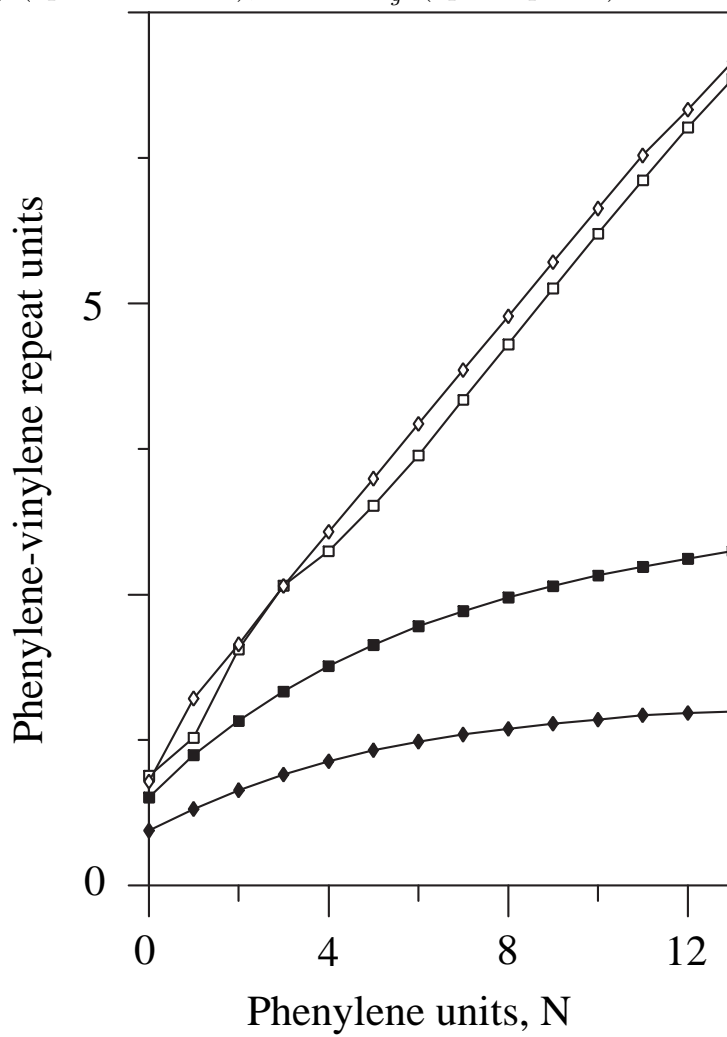}}
\end{figure}

Also shown in Fig. 3 are the lowest two unbound states, with an 
average spacing increasing linearly with oligomer length.   The 
first unbound  state in the $^1\ag$ symmetry sector is denoted 
by $m^1\ag$, where $m=7$ for oligomers of 10 to 13 phenylene units. As shown 
in the next section, this 
is the state with the largest dipole moment with the 
excitonic $1^1\bu$ state, and thus will contribute to the 
non-linear optical spectroscopies \cite{dixit91}.  
In the $^1\bu$ sector we denote the 
first unbound 
state as the $n^1\bu$ state, where $n=4$ for a 13 phenylene unit oligomer. 

The energies of the $n^1\bu$ and $m^1\ag$ states, which
extrapolate to ca.\ 3.2 eV for infinite polymers,
are shown in Fig.\ 2 (a) and (b), respectively. Also shown is
the charge gap, $E_G$, defined as $E_G = E(2N+1)+E(2N-1)-2E(2N)$,  
where $E(2N)$ is the ground state energy of a system with
$2N$ electron.  In the limit of infinite chains this will 
correspond to the energy of an uncorrelated particle-hole pair,
as demonstrated by the fact that its energy extrapolates to ca.\ $3.2$ eV,
in agreement with the energies of the $m^1\ag$ and $n^1\bu$ states
\cite{footnote2}.
Our result for the band threshold agrees with an SCI calculation on a 
re-parametrised P-P-P model by Shimoi and
Abe \cite{shimoi96}.
These considerations of the energetics and mean
separation of the low lying 
states lead us to deduce that there are bands of both
$^1\bu$ and $^1\ag$ excitons below the
conduction band threshold.   

\begin{figure}[htbp]
\caption{
Single-particle weight of the key lowest bound and band
states  as a function of the number of phenylene units.
$1^1\bu$ (solid diamonds), $n^1\bu$ (open diamonds),
$2^1\ag$ (solid squares), $m^1\ag$ (open squares).
}
\centerline{\epsfxsize=10.0cm \epsfbox{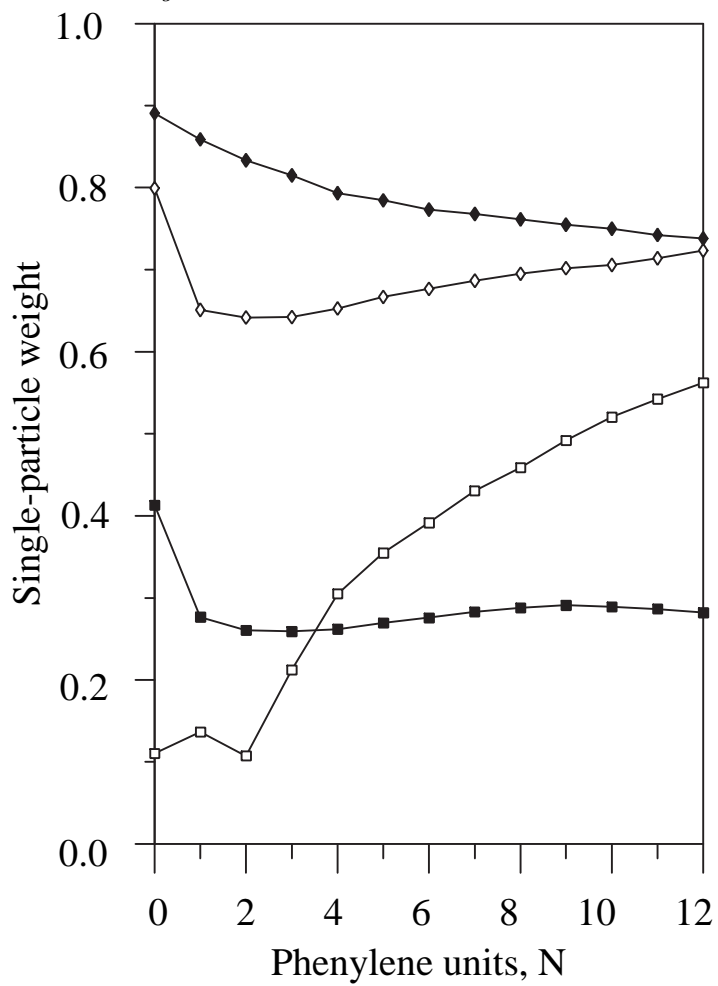}}
\end{figure}

The spatial correlation function,
$C_{ij}(\left|n\right\rangle)$,
may also be used to calculate the total weight
of single-particle excitations in a given excited state compared to
the ground state, defined as:
\begin{eqnarray}
W(\left|n\right\rangle) = \sum_{i\,j} C_{ij}^2(\left|n\right\rangle)
\label{weight}
\end{eqnarray}
Fig.\ 4 shows the single-particle weight for the most important
singlet excited states. For the $^1\bu$ states there is no
substantial difference between the excitonic and unbound states;
both of them have a single-particle weight of about 0.75
for the longest oligomers studied.
In contrast, the $^1\ag$ states show different values
and behaviour of
$W(\left|n\right\rangle)$ as a function of $N$.
The share of the single-particle excitations in the
lowest unbound $m^1\ag$ state increases as the oligomer size increases,
approaching the values of the $^1\bu$ states. For the excitonic
$2^1\ag$ state, however, the single-particle weight remains
approximately constant, at about 0.28, independent of
the system size. The small value of $W(\left|n\right\rangle)$ for
the $2^1\ag$ state
is consistent with the predominately triplet-triplet
character of a strongly correlated state. The same values of 
about 0.1-0.3 were found for other excitonic states of $^1\ag$ symmetry.

\subsection{The Triplet Spectrum}

Photomodulation and photoinduced absorption probes of PPV
have revealed the positions of the low-energy
triplet excitations. The lowest $^3\bu$ state was found at
$1.4$ eV \cite{leng94} and the
triplet-triplet excitation at $1.45$ eV \cite{leng94}
or 1.4 eV \cite{pichler93}, thus giving
the energy of the lowest $^3\ag$ state at 2.80-2.85 eV. 
The results of our calculation
of the lowest triplet excitations are shown in Fig.\ 5. For the system
with 14 phenylene rings the
energy of the  $^3\bu$ state is 1.67 eV, while the
energy of the $^3\ag$ state is 3.41 eV. A 
polynomial fit
gives the energies of the $^3\bu$ and $^3\ag$ states
at 1.64 eV and 3.27 eV, respectively, and 1.63 eV for the transition
between them, in reasonable agreement with experiment.

\begin{figure}[htbp]
\caption{
The energies of the lowest triplet states, $^3\bu$ (solid diamonds)
and $^3\ag$ (solid squares), as a function of the number of
phenylene units.
}
\centerline{\epsfxsize=10.0cm \epsfbox{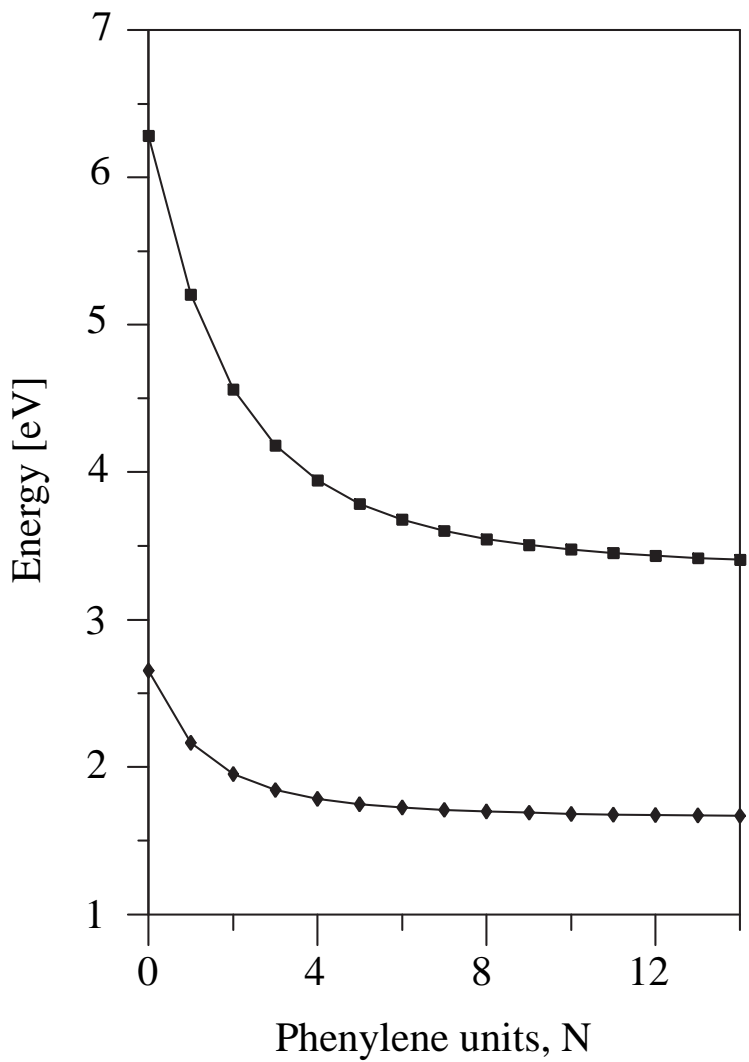}}
\end{figure}

\section{Optical Spectroscopies}

Having calculated the energies, symmetries and spatial 
correlation functions of the low lying states a comparison to 
and interpretation of experiment can be made via the optical spectroscopies. 
The nonlinear optical properties of PPV 
can be related to the third-order 
macroscopic susceptibility $\chi^{(3)}$ which,
in turn, results from the third-order 
microscopic hyperpolarizability $\gamma_{xxxx}$:
\beq
\chi^{(3)}_{xxxx}(-\omega_{\sigma};\omega_1,\omega_2,\omega_3) 
\propto \gamma_{xxxx},
\eeq
where $\omega_{\sigma} = \omega_1 + \omega_2 + \omega_3$.
The calculation 
of $\gamma_{xxxx}$ can be performed using the sum-over-states method 
(see, e.g., \cite{orr71}): 
\bea
\gamma_{xxxx}(-\omega_\sigma;\omega_1,\omega_2,\omega_3) =  
K(-\omega_\sigma;\omega_1,\omega_2,\omega_3) (-\hbar)^{-3} \nonumber \\
I_{1,2,3} ( \sum_{A,B,C} ( \frac{\mu_{gA}\mu_{AB}\mu_{BC}\mu_{Cg}}
{(\omega_A - \omega_{\sigma})(\omega_B - \omega_1 -
 \omega_2)(\omega_C - \omega_1)}+
\frac{\mu_{gA}\mu_{AB}\mu_{BC}\mu_{Cg}}
{(\omega_A^* + \omega_3)(\omega_B - \omega_1 -
 \omega_2)(\omega_C - \omega_1)}+ 
\nonumber \\
\frac{\mu_{gA}\mu_{AB}\mu_{BC}\mu_{Cg}}
{(\omega_A^* + \omega_1)(\omega_B^* + \omega_1 +
 \omega_2)(\omega_C - \omega_3)}+
\frac{\mu_{gA}\mu_{AB}\mu_{BC}\mu_{Cg}}
{(\omega_A^* + \omega_1)(\omega_B^* + \omega_1 + 
\omega_2)(\omega_C^* + \omega_{\sigma})})-
\nonumber \\
\sum_{A,C} ( \frac{\mu_{gA}\mu_{Ag}\mu_{gC}\mu_{Cg}}
{(\omega_A - \omega_{\sigma})(\omega_A - \omega_3)(\omega_C 
- \omega_1)}+ 
\frac{\mu_{gA}\mu_{Ag}\mu_{gC}\mu_{Cg}}
{(\omega_A - \omega_3)(\omega_C^* + \omega_2)(\omega_C - \omega_1)}+
\nonumber \\
\frac{\mu_{gA}\mu_{Ag}\mu_{gC}\mu_{Cg}}
{(\omega_A^* + \omega_{\sigma})(\omega_A^* + \omega_3)(\omega_C^* + 
\omega_1)}+ 
\frac{\mu_{gA}\mu_{Ag}\mu_{gC}\mu_{Cg}}
{(\omega_A^* + \omega_3)(\omega_C - \omega_2)(\omega_C^* + \omega_1)})),
\eea
where $\mu_{ij}$ is the dipole matrix element for the transition between 
the states $i$ and $j$, and $K(-\omega_\sigma;\omega_1,\omega_2,\omega_3)$ 
is a numerical constant which depends on the values of $\omega_{\sigma}$, 
$\omega_1$, $\omega_2$, $\omega_3$ \cite{orr71}.
$I_{1,2,3}$ denotes the average of all terms generated by permuting
$\omega_1$, $\omega_2$ and $\omega_3$.
A finite lifewidth of the levels A, B, C should be 
taken into account in order to calculate $\gamma_{xxxx}$ at the 
resonance points properly. In our calculations, the linewidth
was taken to be 0.04 eV.

The sum in equation (6) is over all states. However, due to the 
fact that the ground state belongs to the $^1A_g^+$ symmetry sector, 
the dipole matrix elements are non-zero only for the transitions 
between $^1A_g^+$  and $^1B^-_{u}$. Thus, the states A and C in (6)
are of $^1B_{u}^-$ symmetry, while the state B, as well as the ground
state, are of $^1A_g^+$ symmetry.

\subsection{Oscillator Strengths}

The results for some of the most important transitions
for oligomers of different sizes
 are given in Table 2. 
The transition with the largest oscillator strength is that between
the ground state and the $1^1\bu$ exciton. Also, there is a
 large oscillator 
strength between the $1^1\bu$ exciton and
the first unbound $^1\ag$ state, $m^1\ag$, as well as between the
$2^1\ag$ exciton and the first unbound $^1\bu$ state, $n^1\bu$.
The transitions between the excitonic, as well as between the
unbound states have substantially smaller oscillator strength.

\begin{table}[htbp]

\centering

\begin{tabular}{|c|c|c|c|c|c|}
\hline $N$ & $1^1\ag \rightarrow 1^1\bu$ & $1^1\bu \rightarrow 2^1\ag$ 
& $1^1\bu \rightarrow m^1\ag$&$2^1\ag \rightarrow n^1\bu$ 
& $m^1\ag \rightarrow n^1\bu$ \\
\hline 0 & 15.942 (4.17) & 0.620 (1.07) & 0.134 (5.34) & 16.644 (3.01) 
& -3.500 (-1.27) \\ 
\hline 1 & 28.326 (3.52) & 0.400 (0.54) & 0.389 (2.68) & 29.284 (2.48) 
& 0.935 (0.33) \\ 
\hline 2 & 39.612 (3.18) & 0.733 (0.38) & 7.985 (2.01) & 40.196 (2.02) 
& 2.771 (0.38) \\ 
\hline 3 & 49.898 (2.99) & 1.225 (0.32) & 49.332 (1.68) & 50.632 (1.68) 
& 10.981 (0.32) \\ 
\hline 4 & 60.021 (2.88) & 1.905 (0.30) & 78.323 (1.51) & 57.173 (1.42) 
& 17.269 (0.22) \\ 
\hline 5 & 69.481 (2.80) & 2.672 (0.30) & 90.479 (1.39) & 61.899 (1.24) 
& 20.306 (0.15) \\ 
\hline 6 & 79.035 (2.76) & 3.579 (0.30) & 93.486 (1.30) & 64.469 (1.09) 
& 19.523 (0.09) \\ 
\hline 7 & 88.383 (2.73) & 4.430 (0.30) & 75.882 (1.23) & 65.876 (0.98) 
& 14.281 (0.05) \\ 
\hline 8 & 97.784 (2.72) & 5.149 (0.31) & 79.097 (1.18) & 66.521 (0.89) 
& 7.760 (0.02) \\ 
\hline 9 & 106.932 (2.70) & 5.795 (0.32) & 73.221 (1.14) & 39.431 (0.84) 
& 7.474 (0.02) \\ 
\hline 10 & 116.580 (2.70) & 6.365 (0.32) & 63.492 (1.10) & 59.735 (0.77) 
& -7.971 (-0.01) \\ 
\hline 11 & 124.450 (2.70) & 6.986 (0.32) & 54.943 (1.07) & 60.744 (0.73) 
& -17.293 (-0.02) \\ 
\hline
\end{tabular}

\caption{Calculated oscillator strengths for selected transitions in 
oligophenylenes with different number of phenylene units, $N$. 
(The corresponding energy differences in eV are shown in brackets.)}

\label{oscstr}

\end{table}

\subsection{Two Photon Absorption}

\begin{figure}[htbp]
\caption{
The calculated TPA spectrum
of a 10 phenylene unit 
oligomer.
}
\centerline{\epsfxsize=10.0cm \epsfbox{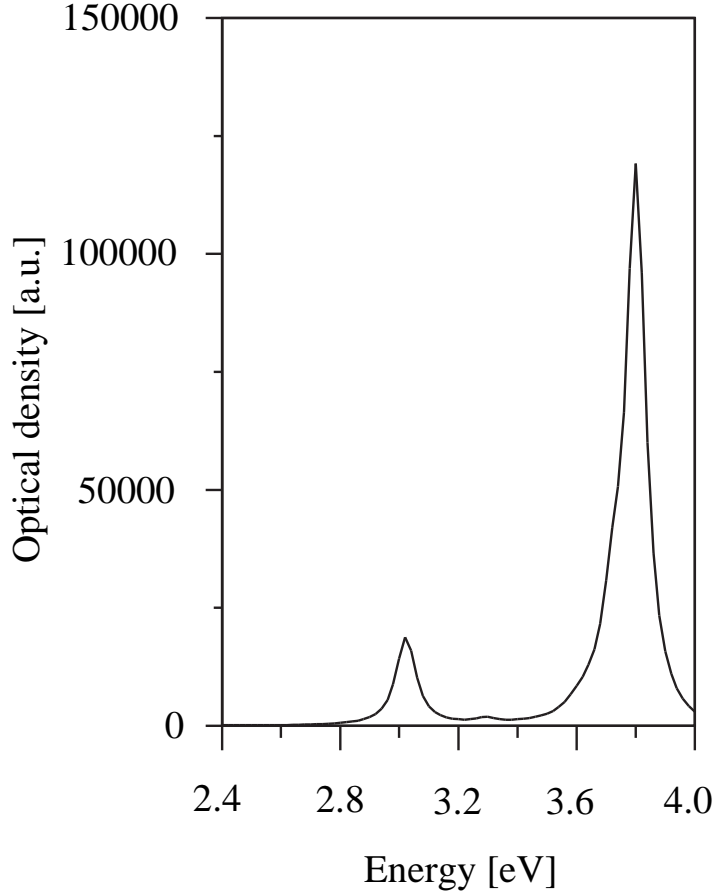}}
\end{figure}

 Fig.\ 6 shows the 
calculated TPA spectrum of a 
10 phenylene unit oligomer.  The weaker peak
at 3.0 eV corresponds to the $2^1\ag$ state, while
the stronger peak at 3.8 eV is the $m^1\ag$ state.
Baker et al.\cite{baker93} performed two photon fluorescence measurements
on
PPV.  They observed a strong signal at 2.9 eV and a weak feature at
3.2 eV.  An interpretation of the relative weights of these peaks is
complicated, because instrumental sensitivity determines the strengths of the 
observed transitions.
Long et al. \cite{long} observed a weak low energy peak at
ca.\ 3.0 eV, and evidence for a steep increase in absorption at ca. 3.2 eV.
This sharp increase in the TPA at 3.2 eV has also been observed
by Meyer et al. \cite{meyer97}
in a derivative of PPV.
We ascribe the low and high energy features observed in \cite{baker93}
and \cite{long}
to the $2^1\ag$ and $m^1\ag$ states,
respectively, while we
interpret the strong 3.3 eV absorption of \cite{meyer97}
as the $m^1\ag$ state.

\subsection{Electroabsorption}

In order to take into account the vibronic structure of the EA spectra 
the sums in (6) are carried out over the vibrational levels 
$\omega_I + n\omega$ (where $\omega_I$
is the electronic state energy, $n$ is an integer and $\omega$ is the 
energy of a phonon). The dipole moments are multiplied by 
the relevant Franck-Condon overlap factor, $F_{pq}$, which is given by
\cite{soos94,mukhopadhyay}
\beq
F_{pq}(a)=\frac{e^{-a^2/4}}{\sqrt{2^{p+q}p!q!}}
\sum_{r}\frac{2^r(-1)^{q-r}a^{p+q-2r}p!q!}{r!(p-r)!(q-r)!},
\eeq
where $p$ and $q$ are the phonon levels between which the transition 
occurs, $a$ is the difference in configurational coordinate between 
the two electronic states involved and the sum is up to the smaller 
of $p$ or $q$.  To simplify the calculation the same phonon energy 
was used for all the states.  

The existence of a range of conjugation lengths within the polymer 
films results in the excited states existing over a range of 
energies.  Further, since the nonlinear response is  strongly 
dependent on the conjugation length, the longer segments within 
the distribution will contribute more to the overall response 
of the system than the shorter segments.  Liess et al. \cite{liess} modeled 
this effect by introducing an asymmetric distribution function 
$\zeta(\omega^{'})$ and calculated the function 
\beq
\chi^{(3)}_{film}(-\omega;0,0,\omega) = \int_{-\delta}^{+\delta}
\zeta(\omega^{'})\chi^{(3)'}_{SOS}(-\omega;0,0,\omega)d\omega{'},
\eeq
where $\omega^{'}$ is the difference between energies of excited
states in finite oligomer and infinite system, and the
$\chi^{(3)'}_{SOS}$ is the sum-over-states susceptibility including 
vibronic effects, calculated with the energies $\omega_A, \omega_B, \omega_C$
increased by $\omega^{'}$. Near to resonances the measured EA 
signal is proportional 
to the imaginary part of the nonlinear susceptibility 
$\chi^{(3)}(-\omega;0,0,\omega)$. 
The lineshape calculated from (8) can thus be compared 
directly to the measured EA lineshapes.

\begin{figure}[htbp]
\caption{
(a) The calculated EA spectrum of a 10 phenylene unit
oligomer. (b) The experimental EA spectrum of a PPV thin 
film, from \protect\cite{martin98}
}
\centerline{\epsfxsize=10.0cm \epsfbox{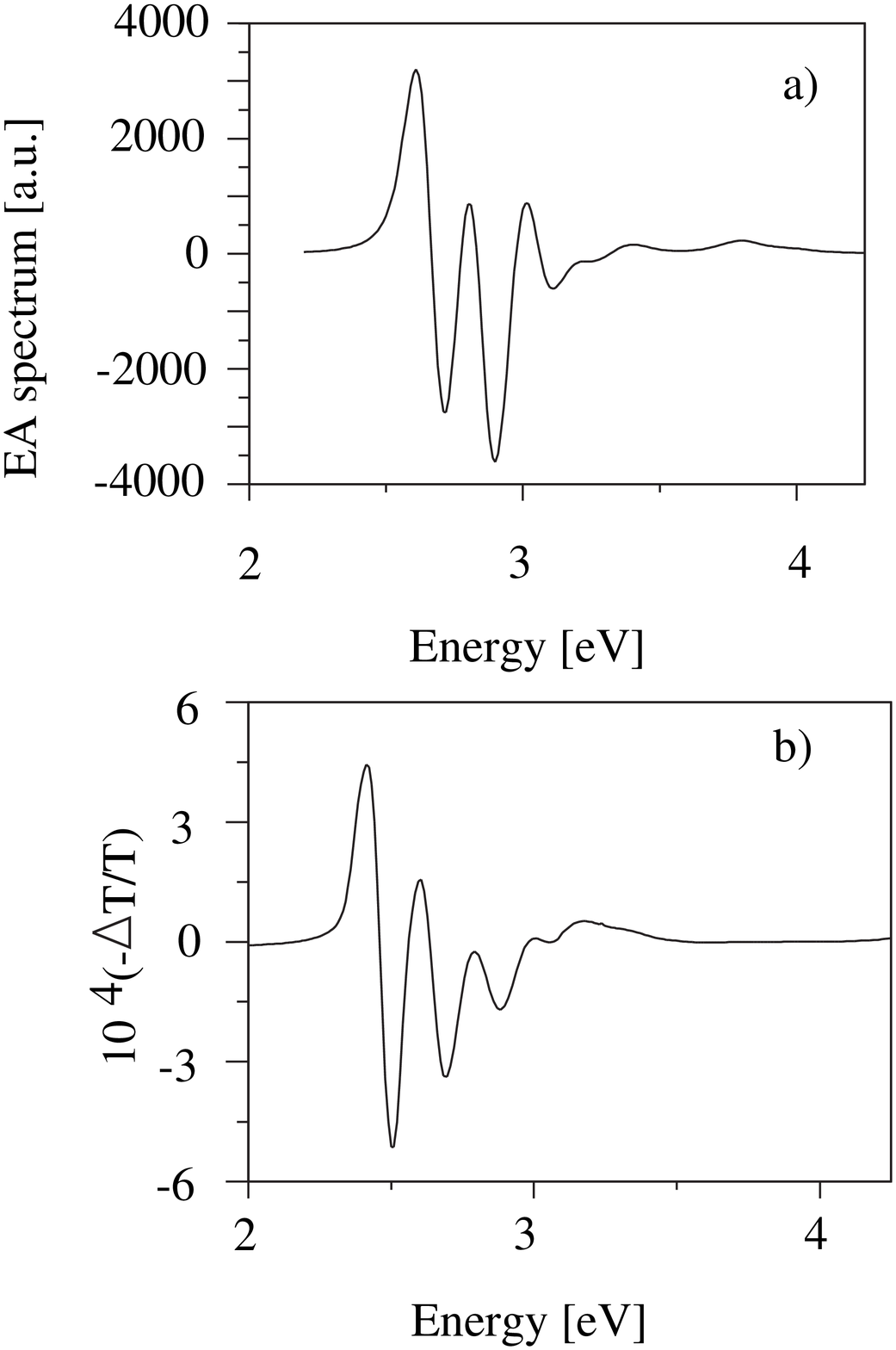}}
\end{figure}

The calculated EA spectrum of a 10 phenylene unit oligomer is 
shown in Fig.\ 7 (a). 
The three sharp peaks are the phonon-split
derivative-like feature corresponding to
the red-shifted $1^1\bu$ exciton, while the   
maximum at ca.\ 3.8 eV corresponds to the conduction 
band threshold 
$\ag$ and $\bu$ states.  The weak 
feature below the  $m^1\ag$ state is the 
$2^1\ag$ state. The experimental EA spectrum from
\cite{martin98} is shown in Fig.\ 
7 (b).  The broad maximum at 3.2 eV, with no correspondence in the derivative 
of the linear absorption, signifies an $^1\ag$
state.  That this
state is the band threshold $m^1\ag$
state is confirmed by
the TPA spectrum discussed above,  the
observation of
 a $^1\bu$ state (namely the
$n^1\bu$ state) at the same energy
by THG \cite{mathy96} and the onset
of photo-conductivity at ca. 3.2 eV \cite{chandross94}.
It is most likely that the $2^1\ag$ state, observed in the TPA, has been 
obscured either by 
the vibronic structure of the $1^1\bu$ exciton, or by the $m^1\ag$ state 
(which is at
a lower energy than the theoretical calculation).

\section{Discussion and Conclusions}

The parametrised 2-MO model of the phenyl based semiconductors
predicts bands of $^1\bu$ (`s'-wave excitons) and $^1\ag$ (`p'-wave
excitons) below the conduction band threshold.  The $1^1\bu$ exciton
energies are in close agreement with oligomer and thin film results.
The experimental two-photon fluorescence \cite{baker93}
and TPA \cite{long}
lends support to there being an excitonic
 $^1\ag$ state at ca.\ 2.9 eV.

The conduction band threshold states have been identified by their
mean particle-hole spacing, and by the fact that their energies approach
the charge gap.  These states have strong transition dipole moments to the
low lying excitonic states and thus contribute strongly to the 
non-linear optical spectroscopies, i.e. TPA and EA.  A comparison of the
calculated EA and TPA spectra with the measured spectra strongly
suggests that the conduction band threshold states lie at ca. 3.2 eV in
PPV thin films.  This result is confirmed by the observation of a $^1\bu$
state by THG, and is consistent with photo-conductivity experiments.

The experimental thin film band 
threshold of $3.2$ eV lies
lower than the theoretical prediction of 3.8 eV for a 10 
phenylene unit oligomer.
The reasons for this are two fold.  First, the 
conjugation length in thin films may be somewhat larger than 10 
phenylene
units, so an energy tending towards the 
infinite polymer result
of 3.2 eV could be expected.  Second, and more importantly, as shown by Moore
and
Yaron \cite{moore97},  the 
theoretical single chain predictions of the band threshold states
have less relevance to thin films than do the predictions of the 
excitonic states, because the band states are 
more subject to the polarisation effects of the surrounding 
medium.   Moore and Yaron considered
 a model system of a solvent polyene chain of
24 atoms and a solute polyene chain of between 2 and 18 carbon atoms, 
separated by 
$4 \AA$.  Solving the P-P-P model, they found that the $\bu $ exciton was 
solvated
by 0.06 eV, while the charge-gap was solvated by 0.38 eV, leading to  a 
reduction in the binding energy of 0.32 eV.  In \cite{moore98} this 
calculation
was extended to 18 solvent chains.  Taking the thermodynamic limit, they
predicted a reduction of the binding energy of 1.3 eV.  Such  a large
reduction
may not be applicable to PPV, but fairly large solvation energies are 
expected.

The interpretation  that the band 
threshold is at $3.2$ eV gives an experimental estimate of the 
energy differences between the vertical energies 
of the $1^1\bu$ and 
$2^1\ag$ excitons  and the band gap as ca. $0.4$ eV and $0.2$ eV, respectively.
The difference in energy between the relaxed $1^1\bu$ state (at 2.4 eV) and the
band gap is 0.8 eV.

Our calculated $1^1\bu$ exciton binding energy, ranging from ca 0.9 eV for a 
13 phenylene unit oligomer to ca.\ 0.6 eV for infinite chains, is almost 
certainly an 
over-estimate of the true binding energy.  The multi-chain solute-solvent 
calculation discussed above \cite{moore97},
\cite{moore98} suggests that it is at least 0.3 eV 
too high.

Shimoi and Abe \cite{shimoi96} and Chandross and Mazumdar \cite{chandross97} 
both predicted a binding energy of 0.9 eV for a single chain.  Again, 
solid-state solvation is expected to decrease this value. 
The phenomenological model by Gartstein, Rice and Conwell
\cite{gartstein95}
gives binding energy between 0.2 and 0.4 eV,
depending on the parameter set. 
The semiconductor band 
calculation of Gomes da Costa and Conwell \cite{gomes93}, which incorporates 
three-dimensional effects, predicted a binding energy of 0.4 eV.
The calculation of Beljonne et al. \cite{beljonne98}  is reasonably
consistent
 with
ours.  They also used a two band model to describe the low energy
excitations. However, the parameters are obtained directly from an INDO
Hamiltonian, and the model is solved for oligomers of up to 6
phenyl(ene) rings
using MRD-CI.  Their predictions
of the $1^1\bu$ at 3.13 eV,
the $2^1\ag$ at 3.78 eV, the $m^1\ag$ at 4.28 eV and the
$n^1\bu$ state at 4.73 eV for an
eight phenylene oligomer, are similiar to ours, although somewhat
blue shifted. They
also 
interpret the $n^1\bu$ state as the band threshold state.

The lowest triplet state is calculated to be at ca.\ 1.6 eV, with the
triplet-triplet gap being ca.\ 1.6 eV.  These predictions are in reasonable
agreement with the experimental results \cite{leng94}, \cite{pichler93}.

It is instructive to compare the results of this calculation to our earlier
calculation on PPP \cite{barford98}
 and to P-P-P model calculations on polyacetylene (PA)
\cite{boman98} \cite{lavrentiev98}.
PPV, being composed of phenylene and vinylene repeat units is, in a sense,
a hybrid of PPP and PA.  The PPP calculation predicts only a band of
$^1\bu$ excitons, with the $2^1\ag$ state representing the band threshold.
In contrast, most calculations predict that in PA the vertical energy
of the $2^1\ag$ state lies below the $1^1\bu$
\cite{lavrentiev98}.  The PPV calculation lies
intermediate to both of these, with both $^1\bu$ and $^1\ag$ excitons.
 Thus, the
vinylene unit behaves as a more highly correlated unit
 than the phenylene unit.

In conclusion, we have presented theoretical
evidence for the existence of
$\bu$ and $\ag$ excitons in PPV. The theoretical calculation was based on the
suitably parametrised two-state model of conjugated semiconductors.
Our interpretation of the experimental non-linear spectroscopies,
in the light of our theoretical calculation, leads to an
estimate of energy difference of ca.\ 0.4 eV between the vertical
energy of the $1^1\bu$ exciton and the band gap and ca.\ 0.8 eV between the
relaxed energy of the $1^1\bu$ exciton and the band gap.
This is consistent
with other experimental predictions
\cite{kersting94}, \cite{marks94}.  

{\bf Acknowledgments}

We thank Dr.\ D.\ Beljonne (Mons),
Prof.\ D.\ D.\ C.\ Bradley (Sheffield),
Dr.\ C.\ Castleton (Grenoble) and Prof.\ D.\ Yaron (Carnegie-Mellon)
for useful discussions.  
M.\ Yu.\ L. and S.\ J.\ M.\ are supported by the EPSRC, grants
GR/K86343 and GR/L84209, respectively.
H.\ D.\ is supported by an EPSRC studentship.
R.J.B.\ is supported by the Australian Research Council.

{\parindent=0pt
{\small{$^*$On leave from Institute of Inorganic Chemistry,
630090 Novosibirsk, Russia}}}



\begin{thebibliography}{40}
\bibitem{footnote} The full P-P-P model calculation of stilbene predicts a 
$^1\ag$ state at 4.39 eV.  However, this is derived from the $B_{3g}$ state, 
and is not described by the 2-MO model \cite{castleton}.
\bibitem{footnote2} We believe that the exact agreement between 
the infinite single chain result and the experimental data
on thin films of shorter
conjugation length (see \S 4) for the band threshold energy is largely
coincidental.
\end{thebibliography}
\end{document}